\documentstyle[twoside,fleqn,hooperstyle,epsfig]{article}

\def\bold#1{\setbox0=\hbox{$#1$}%
     \kern-.025em\copy0\kern-\wd0
     \kern.05em\copy0\kern-\wd0
     \kern-.025em\raise.0433em\box0 }

\newcommand{\gev}{\,\mbox{GeV}}

\def\ga{\mathrel{\raise.3ex\hbox{$>$\kern-.75em\lower1ex\hbox{$\sim$}}}}
\def\la{\mathrel{\raise.3ex\hbox{$<$\kern-.75em\lower1ex\hbox{$\sim$}}}}

\def\m12{m_{1\!/2}}

\begin{document}
\title{Indirect Detection of Neutralino Dark Matter up to TeV Scale}

\author{Dan Hooper
\address{Department of Physics, University of Wisconsin,
    Madison, WI~53706, USA}}

\begin{abstract}
In this paper, we will describe the results of SUSY parameter space searches including minimal supergravity, non-universal supergravity and minimal supersymmetry and the implications on the indirect detection of neutralino dark matter.  We give special attention to the effects of detector thresholds, solar absorption of neutrinos and hadronization of neutralino annihilation products.  These effects are known to be important in calculating accurate event rates \cite{Berg00}.

We chose also to focus on models which predict a heavy lightest neutralino (several hundred GeV to several TeV).  These models have been selected for several reasons including their inaccessibility in future collider searches.  
\end{abstract}

\maketitle

\section{INTRODUCTION}

There has been a great deal of evidence mounting recently for the existence of dark matter consisting primarily of the the lightest supersymmetric particle (LSP) \cite{Jung96}.  In most models, the LSP is the neutralino, a superposition of the supersymmetric partners of the photon, $Z$ boson and neutral Higgs bosons.  By virtue of $R$-parity conservation, the LSP is totally stable and would freeze out during early stages of the universe.  In many models, the predicted relic density of neutralinos is in the cosmologically interesting region \begin{math}  (.05<\Omega_\chi h^2<.3)\end{math} which is further evidence for a significant neutralino dark matter component.

In this paper, we consider a large area of minimal supergravity (mSUGRA), non-universal supergravity and minimal supersymmetry (MSSM) parameter spaces.  We determine which regions predict an acceptable cold dark matter particle candidate and calculate the upward going muon fluxes in neutrino telescopes from neutralino annihilation in the sun and compare these results with direct detection rates for these models.  We have also calculated the rates from annihilation in the Earth, but do not discuss these results here as they are generally negligible compared to the solar rates.

The results presented in this paper are described in more detail in Reference \cite{Halz01}.

\section{BACK-OF-THE-ENVELOPE CALCULATIONS}

In this section, we roughly calculate the event rates for indirect and direct WIMP detection experiments in a model independent fashion.  No references to any specific supersymmetry or other model are required for the following calculations.  These calculations closely follow Reference \cite{Halz95}.

\subsection{Assumptions}

Before starting the calculations, some basic assumptions need to be stated.  First, we assume that the WIMP in question is the major constituent of the measured dark matter halo density: 

\begin{equation}
           \phi_{\chi}=n_{\chi}v_{\chi}=\frac{.4}{m_{\chi}}\frac{\gev}{{\rm{cm}}^{3}}\;3\times 10^{6}\; \frac{{\rm{cm}}}{{\rm{s}}}
\end{equation}

We will also approximate, using dimensional analysis, the cross section for WIMP-Nucleon interactions:

\begin{eqnarray}
	\sigma_{DA} & \equiv & \sigma(\chi N)=(G_{F}m_{N}^{2})^2 \frac{1}{m_{W}^{2}} \nonumber \\
                    & =      & 6 \times 10^{-42} \rm{cm}^2
\end{eqnarray}

Thirdly, we will assume that WIMPs annihilate to neutrinos $\sim$10 percent of the time.  This is simply a reflection of the branching ratios of heavy quarks or W boson pairs to neutrinos.

\subsection{Indirect Detection}

Indirect detection experiments are based on the idea that heavy particles, such as WIMPs will become gravitationally bound in objects such as the Sun and Earth.  Over a time scale generally less than the age of the solar system, these bodies can obtain WIMP densities which will produce an observable flux of annihilation products (generally neutrinos) for a detector on Earth.  

With the assumptions stated and out of the way, we will begin the calculation of indirect event rates from solar WIMP annihilation.  First, the solar gravitational capture cross section is given by:

\begin{equation}
           \sigma_{\rm{sun}}=\rm{f} (1.2 \times 10^{57}) \; \sigma_{DA}
\end{equation}

Where f is the focusing factor and is equal to the ratio of kinetic to potential energy of the WIMP near the Sun and is generally of the order of 10.  The flux of WIMP annihilation products from the Sun in units of number of WIMPs is:

\begin{equation}
        \phi_{\rm{sun}}=\phi_{\chi}\sigma_{\rm{sun}}/4\pi d^{2}
\end{equation}

Where d is the Earth-Sun distance.  Then, using the assumption of ten percent branching ratios of WIMPs to neutrinos, the resulting neutrino flux at Earth is given by:

\begin{equation}
 \phi_{\nu}=.10 \times \phi_{\rm{sun}} = \frac{3 \times 10^{-5}}{m_{\chi}/\gev}\;\rm{cm}^{-2} \rm{s}^{-1}
\end{equation}

The probability to detect a neutrino in an ice-based neutrino telescope is proportional to the neutrino interaction cross section and the muon range within the detector.  This, along with the assumption that the neutrino energy is half (or one third in the case of heavy quark decay) of the WIMP mass, gives us the probability of detecting a neutrino as:
  
\begin{equation}
           P = N_{A} \sigma_{\nu} R_{\mu}
\end{equation}
\begin{equation}
	   \sigma_{\nu}=0.5 \times 10^{-38} (E_{\nu}/\gev)\; \rm{cm}^{2}
\end{equation}
\begin{equation}
	   R_{\mu}$=muon range=$\;500 \;\rm{cm} \; (E_{\mu}/\gev)
\end{equation}
\begin{equation}	  
	   \rightarrow P=2 \times 10^{-13} ({m_{\chi}/\gev})^{2}   
\end{equation}

Thus, the indirect event rate is simply:

\begin{equation}
           R_{\rm{Indirect}}=\phi_{\nu} P = 1.8 \; (m_{\chi}/\gev)\; \rm{yr}^{-1} \rm{km}^{-2}
\end{equation}

This calculation is very approximate in nature and is only intended to provide a rough idea of the factors involved in predicting such an event rate.  Some of the neglected factors include the solar absorption of neutrinos, the hadronization of quarks, neutrinos from other annihilation channels, detector threshold effects, resonances, model-dependent characteristics of the WIMP, coannihilations and coherent nuclear enhancements.  Later in this paper, most of these factors will be considered and included in the results given.

\subsection{Direct Detection}

For the purpose of comparison, we will now proceed to calculate, in a similar manner, the event rates for direct detection experiments.  Direct experiments measure the energy transferred from halo WIMPs directly to the experimental apparatus.  The event rate for such an experiment is given by:

\begin{equation}
           R_{\rm{Direct}}=\frac{1}{m_{N}}\phi_{\chi}\sigma_{DA}=\frac{1.4}{m_{\chi}/\gev}\; \rm{kg}^{-1} \rm{yr}^{-1}  
\end{equation}

This result provides a rough estimate of such an event rate.  It neglects, among other factors, the model-dependent characteristics of the WIMP, coherent nuclear enhancements and the properties and composition of the detector.

Comparing these approximate results for indirect and direct rates, we can gauge the relative effectiveness of such experiments.  For a 10,000 square meter indirect detector and a one kilogram direct detector, direct detectors observe more events only for the case of an extremely light WIMP (less than 10 GeV) which has previously been ruled out \cite{Junk01}.  This result is not general due to the factors neglected in the above calculations.  A much more detailed comparison will be given later in this paper.

In addition to signal comparison, the background for indirect and direct detection experiments should be considered.  Indirect detection background is from atmospheric neutrinos and is determined by their known energy spectrum.  The number of events can vary from a few thousand background events per year per solar pixel for neutrinos of energy around 10 GeV to less than one background event per year per solar pixel for neutrinos of energy near 1 TeV.  Direct detector backgrounds are primarily energy independent and are generally around 300 events per year per kilogram of detector material.  Therefore, at high energies, indirect detectors have background advantages while at low energies, direct detectors have fewer background events.
\begin{figure*}[t]
\begin{center}
  \epsfig{file=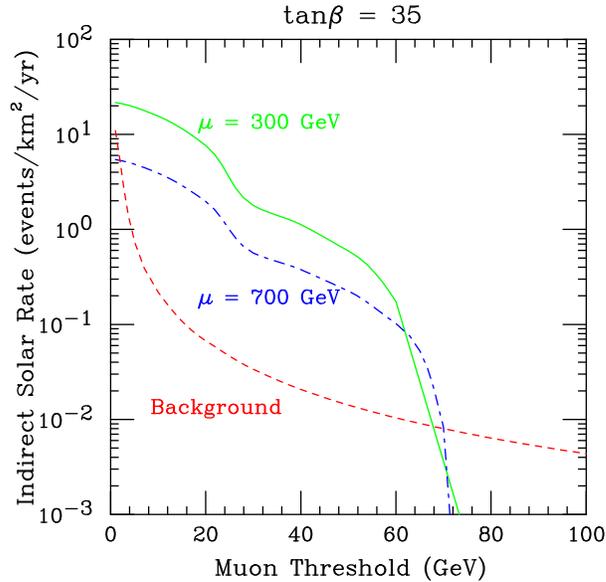,width=8cm} 
\end{center}
\vspace*{-13mm}
\begin{center}
  \caption{Effects of solar absorption, hadronization and threshold on indirect event rates.  The models considered in the figure are the MSSM with $M_{2}=400$ GeV, $\tan \beta=35$ and reflect typical behavior for most models with a similar mass.  Heavier models are effected much less by such factors.}
\end{center}
\end{figure*} 

\begin{figure*}[t]
\begin{center}
  \epsfig{file=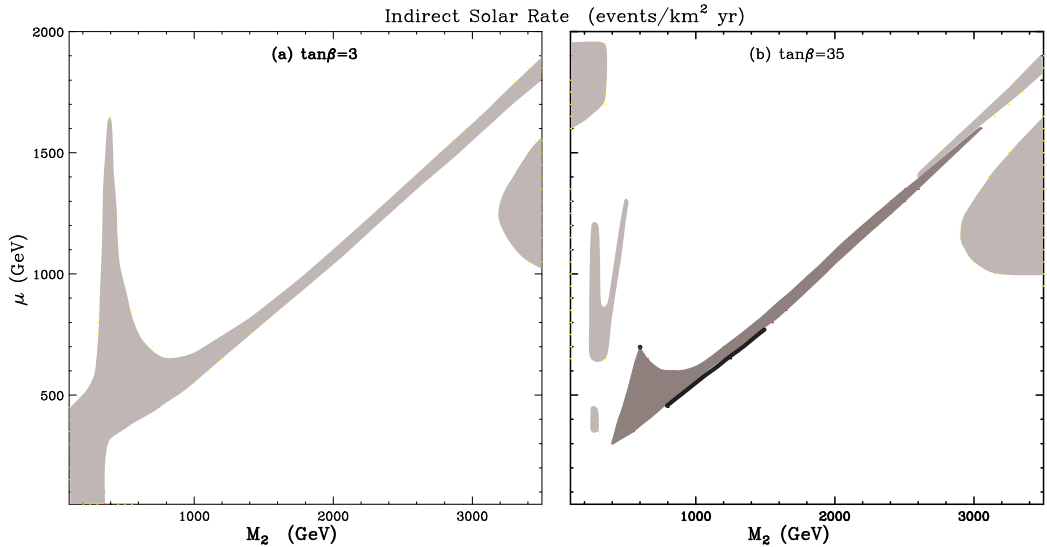,width=14cm} 
\end{center}
\vspace*{-13mm}
\begin{center}
  \caption{Indirect detection rate of neutrinos from solar WIMP-WIMP annihilation in the MSSM.  Regions represent more than 100, more than 10 and less than 10 events per year per square kilometer effective area (darkest to lightest).}
\end{center} 
\end{figure*}

\section{Effects of Hadronization, Solar Absorption and Detector Threshold on Indirect Detection Rates}

To remove as much of the uncertainty in the above calculation as possible, we have taken steps to include many additional factors in our parameter space searches.  To do so, we first consider the dominant annihilation modes for a neutralino WIMP.  For a gaugino-like neutralino, the dominant annihilation modes are $b\bar{b}$ or $\tau \bar{ \tau }$ for $m_{\chi}<m_{W^{\pm}}$ and $b\bar{b}$ for $m_{\chi}>m_{W^{\pm}}$.  For a higgsino-like neutralino, the dominant modes are $Z Z$ or $W^{+}W^{-}$ for $m_{\chi}<m_{t}$ and $t\bar{t}$ for $m_{\chi}>m_{t}$ \cite{Jung95}.  Any quark products, and the resulting neutrino products, lose a fraction of their energy to hadronization as determined by the quark flavor \cite{Jung95}.   

Furthermore, neutrinos produced in the Sun may be absorbed before they escape the solar medium.  We have used the Ritz and Seckel parameterization up to energies around 1 TeV for all neutrino producing annihilation modes \cite{Ritz88}.  Above the TeV scale, corrections were needed and were done in a manner similar to work by Edsjo \cite{Edsj97}.

Indirect detectors generally have neutrino energy thresholds of a few tens of GeV.  We have chosen 25 GeV for the results in this paper.  We have also made the calculations for a threshold of 100 GeV but in the interest of space have neglected these results.  For a more detailed description of these effects, see References \cite{Halz01,Berg00}.   

The total effect of hadronization, solar absorption and threshold can vary with energy from the total destruction of the neutrino signal for masses around 100 GeV to merely a reduction by a factor of two or less in neutrino signal at TeV scale masses.  Figure 1 shows the results of these effects for a typical SUSY model with a light neutralino of about 180 GeV.  The effects are far less significant for heavier models.

The calculations presented in this discussion neglect the effects of coannihilations.  These effects are presented in Reference \cite{Elli99}.

\begin{figure*}[t]
\begin{center}
  \epsfig{file=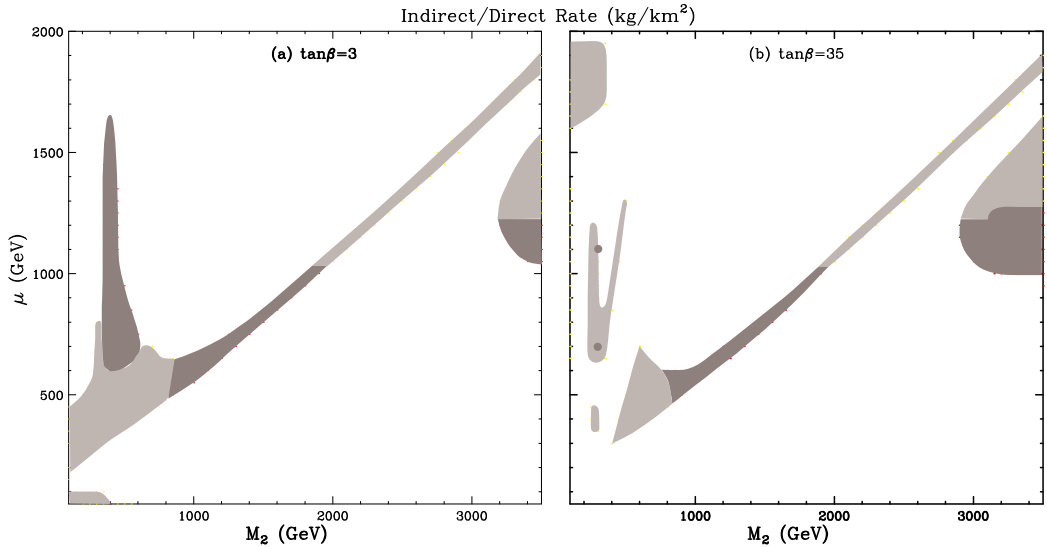,width=14cm} 
\end{center}
\vspace*{-13mm}
\begin{center}
  \caption{Ratio of solar indirect detection rate to direct detection of galactic halo WIMPs in the MSSM.  Regions represent more than 100, more than 10 and less than 10 in units of events per year per square kilometer effective area and events per year per kilogram of detector material (darkest to lightest).}
\end{center}
\end{figure*}

\begin{figure*}[t]
\begin{center}
  \epsfig{file=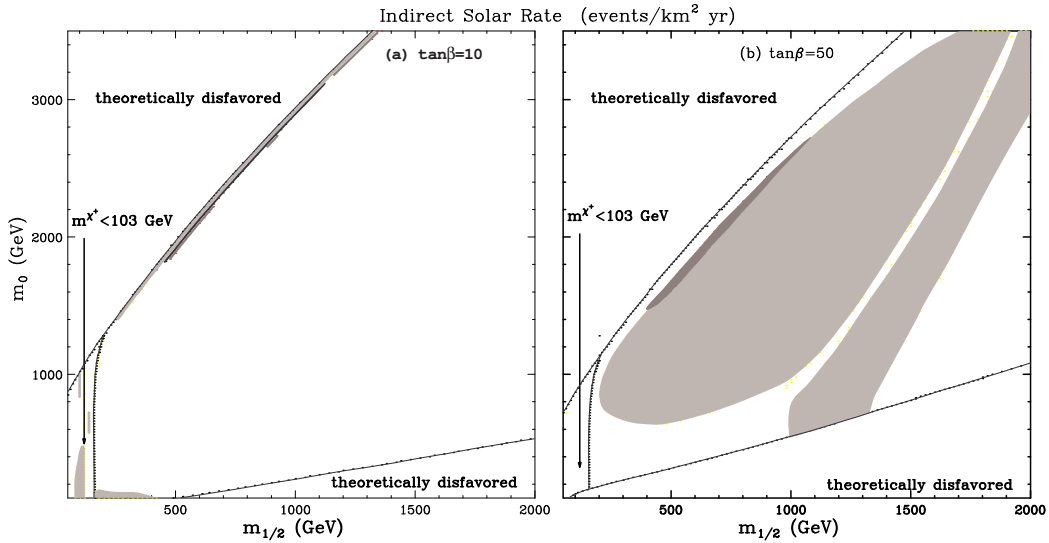,width=14cm} 
\end{center}
\vspace*{-13mm}
\begin{center}
  \caption{Indirect detection rate of neutrinos from solar WIMP-WIMP annihilation in mSUGRA.  Regions represent more than 10, more than 1 and less than 1 in units of events per year per square kilometer effective area (darkest to lightest).  Theoretically disfavored regions are discussed in the text.}
\end{center}
\end{figure*}

\begin{figure*}[t]
\begin{center}
  \epsfig{file=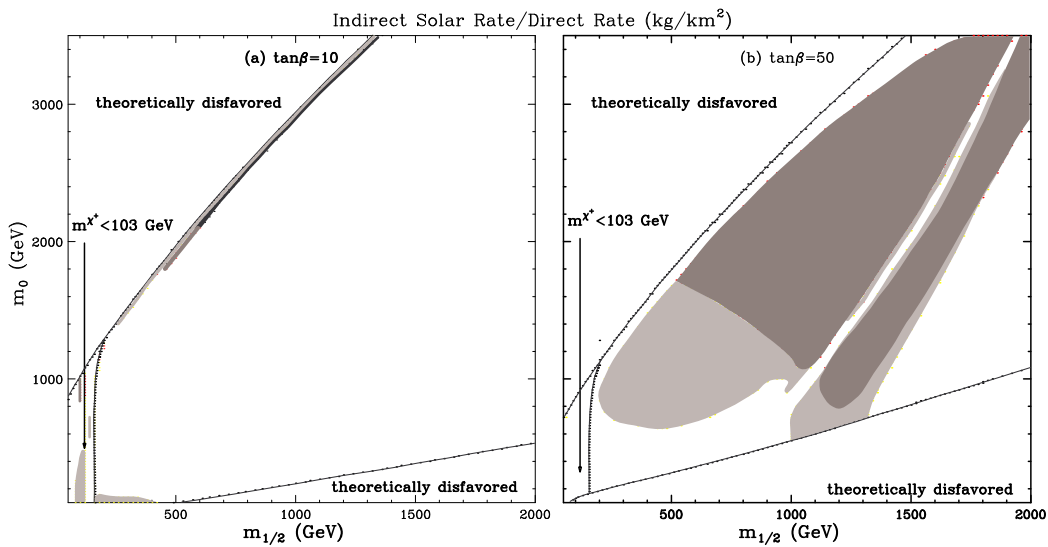,width=14cm} 
\end{center}
\vspace*{-13mm}
\begin{center}
  \caption{Ratio of solar indirect detection rate to direct detection of galactic halo WIMPs in mSUGRA.  Regions represent more than 10, more than 1 and less than 1 in units of events per year per square kilometer effective area and events per year per kilogram of detector material (darkest to lightest).  Theoretically disfavored regions are discussed in the text.}
\end{center} 
\end{figure*}

\begin{figure*}[t]
\begin{center}
  \epsfig{file=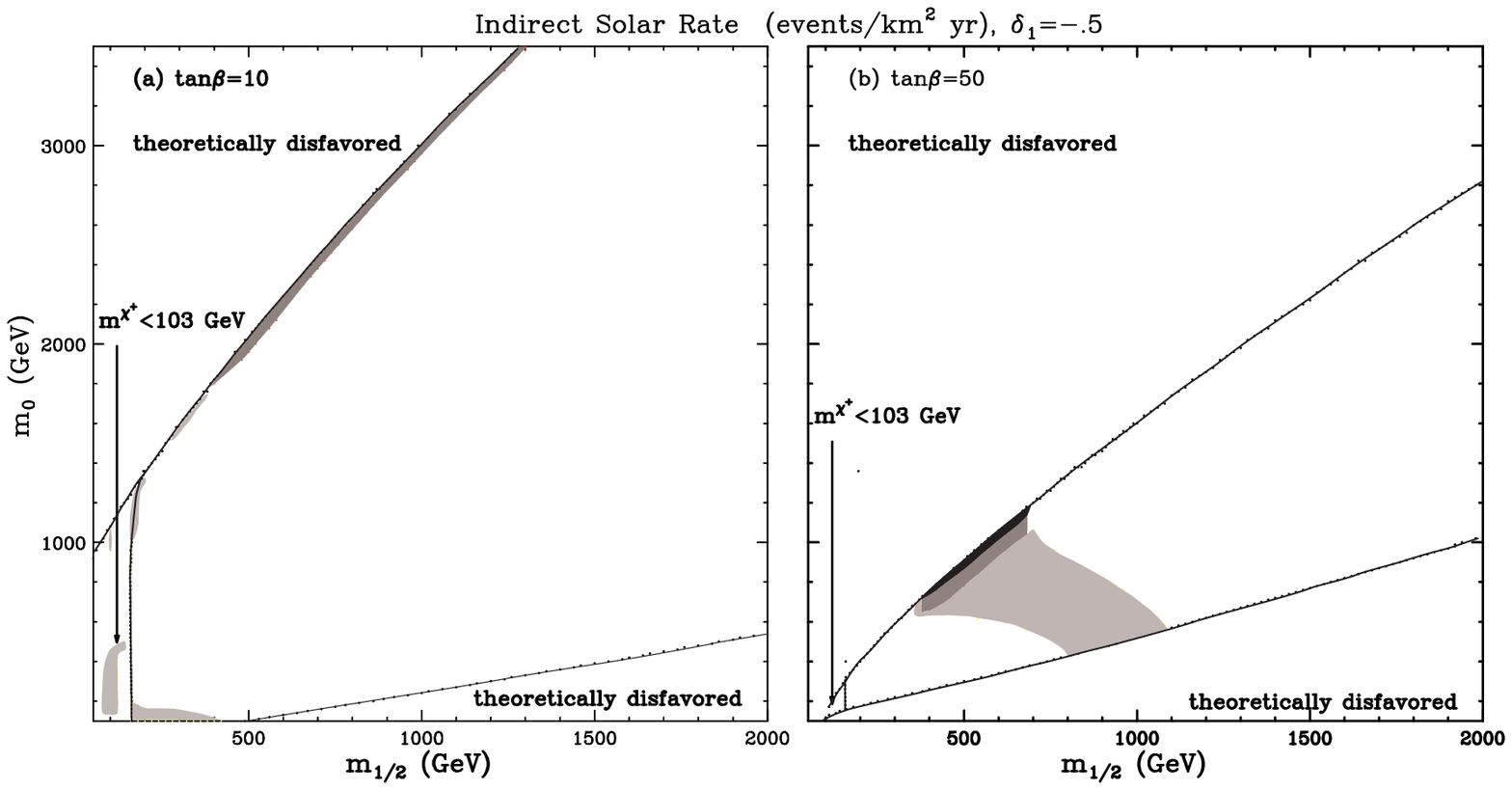,width=14cm} 
\end{center}
\vspace*{-13mm}
\begin{center}
  \caption{Indirect detection rate of neutrinos from solar WIMP-WIMP annihilation in the non-universal SUGRA case of $\delta_{1}=-.5$.  Regions represent more than 10, more than 1 and less than 1 in units of events per year per square kilometer effective area (darkest to lightest).  Theoretically disfavored regions are discussed in the text.}
\end{center}

\begin{center}
  \epsfig{file=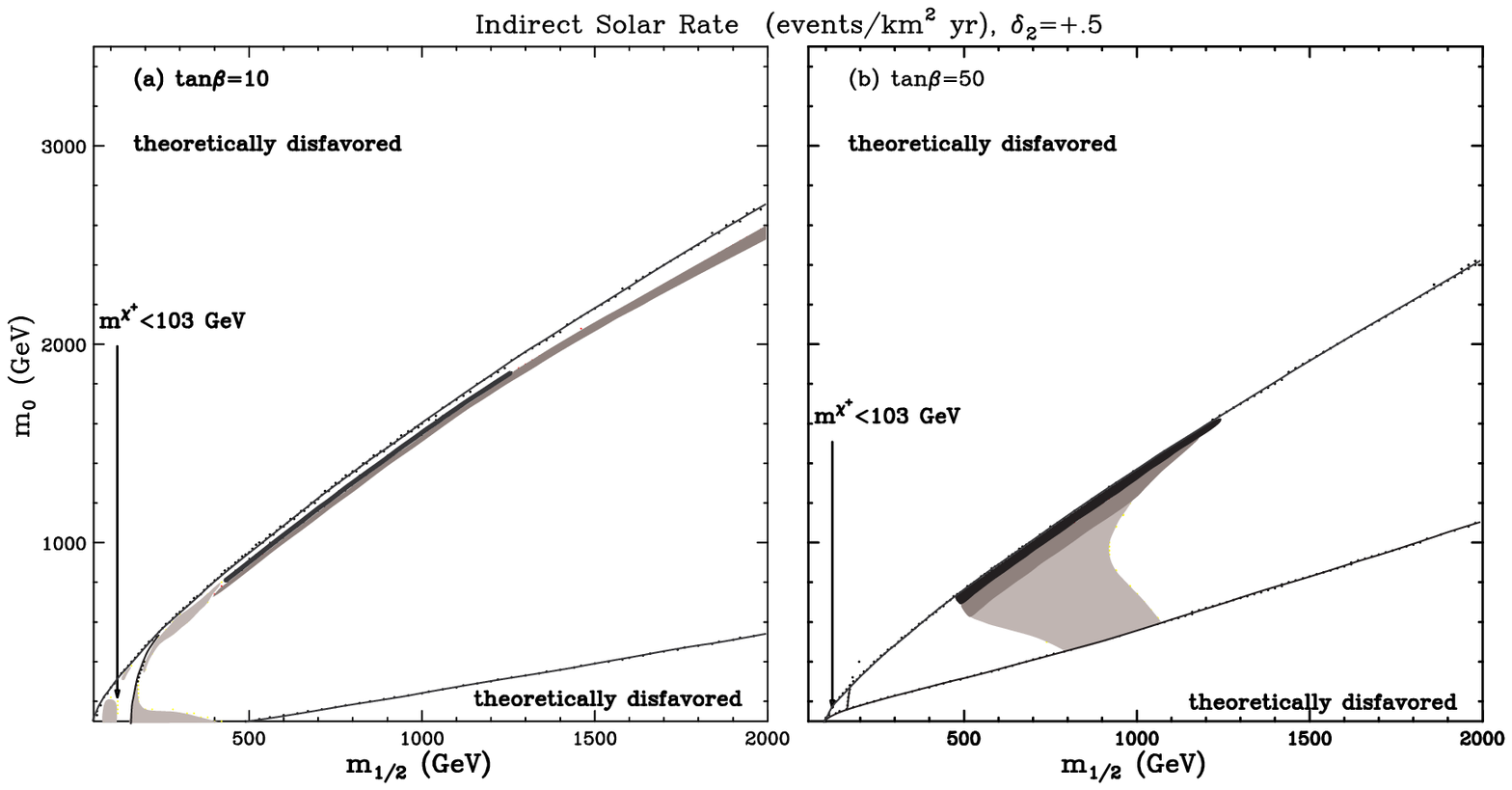,width=14cm} 
\end{center}
\vspace*{-13mm}
\begin{center}
  \caption{Indirect detection rate of neutrinos from solar WIMP-WIMP annihilation in the non-universal SUGRA case of $\delta_{2}=.5$.  Regions represent more than 10, more than 1 and less than 1 in units of events per year per square kilometer effective area and events per year per kilogram of detector material (darkest to lightest).  Theoretically disfavored regions are discussed in the text.}
\end{center}
\end{figure*}

\begin{figure*}[t]
\begin{center}
  \epsfig{file=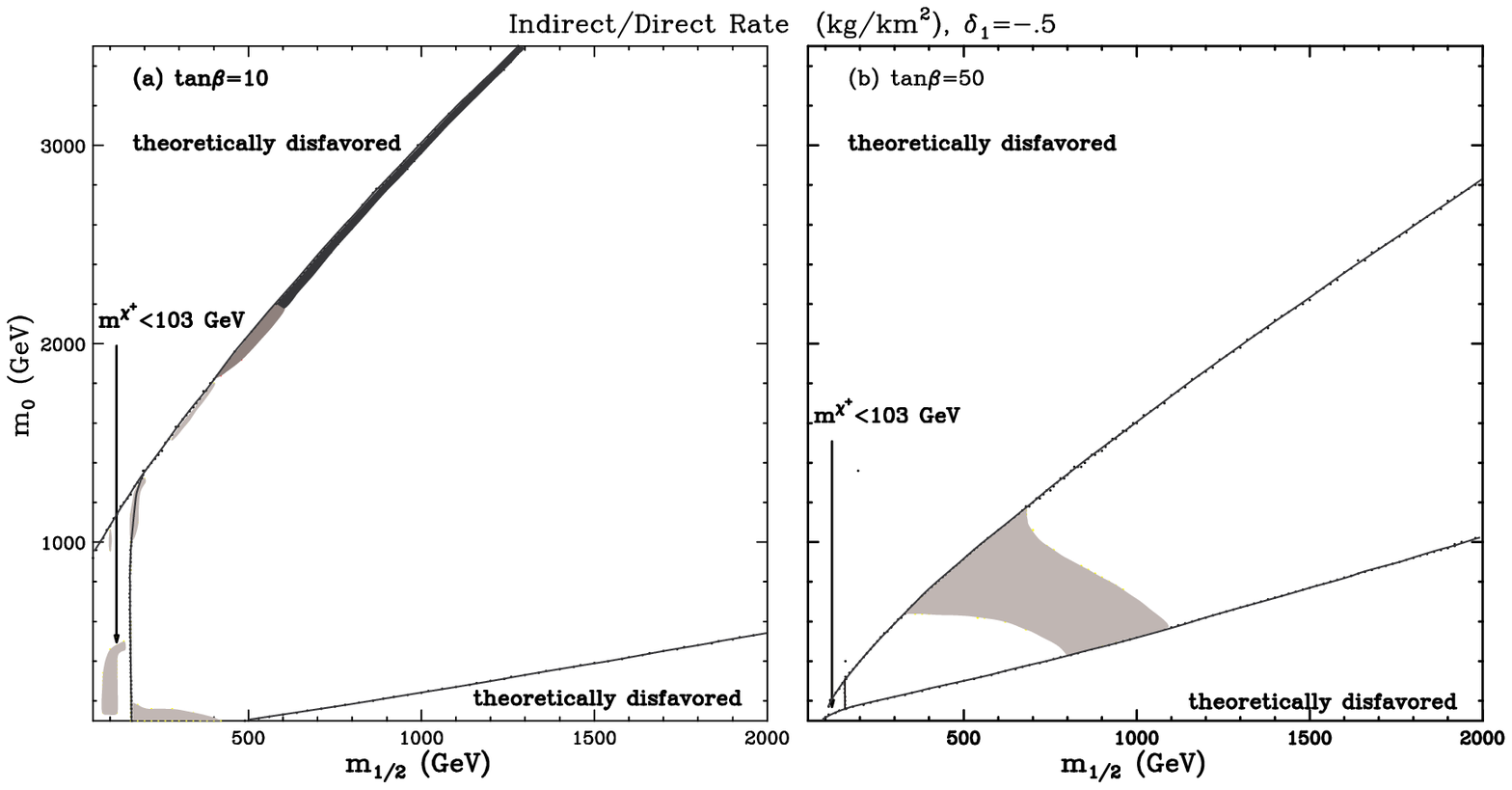,width=14cm} 
\end{center}
\vspace*{-13mm}
\begin{center}
  \caption{Ratio of solar indirect detection rate to direct detection of galactic halo WIMPs in the non-universal SUGRA case of $\delta_{1}=-.5$.  Regions represent more than 10, more than 1 and less than 1 in units of events per year per square kilometer effective area (darkest to lightest).  Theoretically disfavored regions are discussed in the text.}
\end{center}

\begin{center}
  \epsfig{file=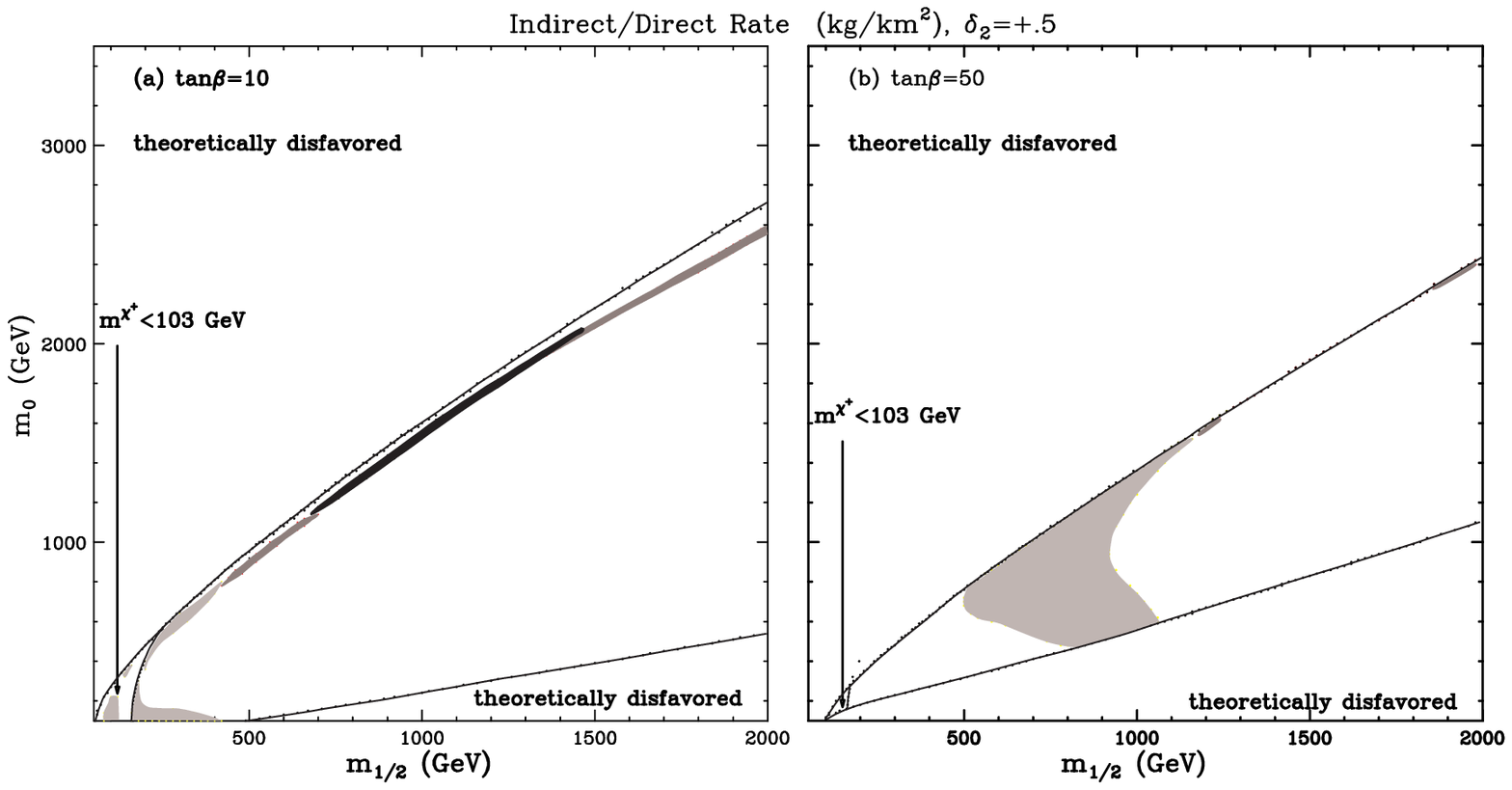,width=14cm} 
\end{center}
\vspace*{-13mm}
\begin{center}
  \caption{Ratio of solar indirect detection rate to direct detection of galactic halo WIMPs in the non-universal SUGRA case of $\delta_{2}=.5$.  Regions represent more than 10, more than 1 and less than 1 in units of events per year per square kilometer effective area and events per year per kilogram of detector material (darkest to lightest).  Theoretically disfavored regions are discussed in the text.}
\end{center}
\end{figure*}

\section{Searches of Supersymmetric Parameter Space}

Beginning with the most simple set of SUSY models, the MSSM, we attempt to identify the model-dependent properties of neutralino dark matter detection.  To simplify the parameter space, we set all squark and slepton masses to 300 GeV or 1.5 times the neutralino mass (whichever is larger) \cite{Halz92}.  We have also made calculations for the choice of 1500 GeV or 1.5 times the neutralino mass but in the interest of space do not display them here.  We then search by varying $\mu$, $M_2$, $\tan \beta$ and the couplings.  We set $M_1$ relative to $M_2$ using the GUT scale relations and we hold $M_{H_2} \simeq 120$ GeV (within 15 GeV).  We consider only those choices of parameters which yield a neutralino with a cosmologically interesting relic density ($.05<\Omega_\chi h^2<.3$) which is the lightest supersymmetric particle (LSP).  Figures 2 and 3 show the indirect detection rate and ratio of indirect to direct detection rates for the MSSM.  Note that when the lightest neutralino is largely gaugino, which is most often the case, the GUT relations cause $m_{\chi}$ to scale with $M_{2}$.

Extending the MSSM to minimal supergravity (mSUGRA), we simplify our model parameters to the universal scalar mass at the GUT scale $m_0$, the universal gaugino mass the GUT scale $m_{1/2}$, the ratio of vacuum expectation values of Higgs scalars $\tan \beta$, the universal trilinear coupling at the GUT scale $A_0$ and the sign of the superpotential term in the Lagrangian $\mu$.  Results shown here are for models with $A_0$=0 and $\mu>0$.  We have calculated results for other models and found generally no significant differences.  Changes in $A_0$ mostly effect masses of third generation superpartners and, therefore, have little effect on the relic density or relevant cross sections \cite{Baer97}.  Figures 4 and 5 show the indirect detection rate and ratio of indirect to direct detection rates in mSUGRA.  

Extending supergravity to the cases of non-universality, we introduce two new parameters $\delta_{1}$ and $\delta_{2}$ which parameterize the divergence of the Higgs masses from the universal scalar mass at the GUT scale: $H_{1}=(1+\delta_{1})m_{0}, H_{2}=(1+\delta_{2})m_{0}$ \cite{Bere96}.  The case of $\delta_{1}=\delta_{2}=0$ reduces to mSUGRA.  In both minimal and non-universal SUGRA, we consider only models which allow for acceptable electroweak symmetry breaking and are tachyon-free in addition to the requirements imposed above in the MSSM models.   Figures 6 through 9 show the indirect detection rate and ratio of indirect to direct detection rates for the non-universal SUGRA.  Note that the mass of the lightest neutralino scales with $m_{1/2}$ and is roughly 40$\%$ of $m_{1/2}$ for a largely gaugino lightest neutralino, which is the case for most mSUGRA and non-universal SUGRA models \cite{Dree95,Care94,Barg97}.

\section{Discussion}

Figures 2 and 3 show the indirect detection rates and the ratio of indirect to direct detection rates for the MSSM.  The regions represent more than 100, more than 10 and less than 10 (in units shown on the figures).  Empty regions represent models with non-cosmologically interesting neutralino relic densities.

The rates are clearly more favorable for $\tan \beta=35$ than for  $\tan \beta=3$ models.  It is difficult to say how many events are necessary for a positive dark matter detection.  A fair estimate of the number needed for a positive result is probably around 10 events per year for a model with a lightest neutralino with mass around a few hundred GeV.  With this in mind, much of MSSM parameter space should be experimentally accessible to kilometer scale neutrino telescopes.  From Figure 3, it is clear that indirect detection is more favorable than direct detection in the vast majority of MSSM models.  This advantage is even greater, in all but the lightest models, if the effects of background are included.

The searches of mSUGRA and non-universal SUGRA space are somewhat less favorable than the MSSM but still show promise.  The regions in Figures 4 through 9 represent more than 10, more than 1 and less than 1 (in units shown on the figures).  Empty regions are either disfavored by their neutralino relic density or by other theoretical constraints such as acceptable electroweak symmetry breaking, the existance of tachyons.    

Much of the SUGRA parameter space produces a few events per year for a kilometer scale neutrino telescope.  This would likely require many years of observation to observe a positive result.  Large portions of the $\tan \beta=50$ region would not be observable even with many years of observation. 
  
\section{Conclusions}
Several important conclusions can be drawn from the results presented in this paper.  First, in most SUSY models, indirect detection is favored over direct detection methods, especially in heavier neutralino models.  Second, the effects of hadronization, solar absorption and detector threshold are significant and should not be neglected.  Third, a significant portion of supersymmetric parameter space, especially within the MSSM should be accessible to present or next generation indirect detection experiments.  More importantly, however, is that the areas of SUSY parameter space which are generally most accessible to indirect detection are the same areas which will be very difficult to probe in collider experiments \cite{Abel00,Abdu99,Acco98}.  With this in mind, if nature has selected a heavy scale for supersymmetry, indirect dark matter experiments may make the first observations of supersymmetry.

\section{Acknowledgements}
Thanks to Chung Kao, Francis Halzen and Vernon Barger who contributed greatly to this work.  Also thanks to Toby Falk and Tilman Plehn for helpful discussions.


\begin{thebibliography}{9}

\bibitem{Berg00} L. Bergstrom, Rept.\ Prog.\ Phys.\ {\bf 63}, 793 (2000).

\bibitem{Jung96} G. Jungman, M. Kamionkowski and K. Griest, Phys. Rep. 267 (5,6) 195-367 (1996).

\bibitem{Halz01} V. Barger, F. Halzen, D. Hooper and C. Kao, MADPH-00-1195, {\it to be published.}

\bibitem{Halz95} F. Halzen, {\it Prepared for International Symposium on Particle Theory and Phenomenology, Iowa State University, 22-24 May 1995}, astro-ph/950820.

\bibitem{Junk01} T. Junk {\it et al.}  [LEP Collaborations], hep-ex/0101015.

\bibitem{Jung95} G. Jungman and M. Kamionkowski, Phys.\ Rev.\ D {\bf 51}, 328 (1995).

\bibitem{Ritz88} S. Ritz and D. Seckel, Nucl.\ Phys.\ {\bf B304}, 877 (1988).

\bibitem{Edsj97} J. Edsjo, Nucl. Phys. B 43 (1995) 265; Ph.D. Thesis, Uppsala University (1993).

\bibitem{Elli99} J. Ellis, T. Falk, K.A. Olive and M. Srednicki, Astropart.\ Phys.\ {\bf 13}, 181 (2000).

\bibitem{Baer97} H. Baer and M. Brhlik, Phys. Rev. D 57 (1997) 567.

\bibitem{Halz92} F. Halzen, M. Kamionkowski and T. Stelzer, Phys. Rev. D 45 (1992) 4439.
 
\bibitem{Bere96} G. Mignola, V. Berezinsky, A. Bottino, J. Ellis, N. Fornengo and S. Scopel, Nucl.\ Phys.\ Proc.\ Suppl.\ {\bf 48}, 50 (1996).

\bibitem{Dree95} M. Drees and S.P. Martin, hep-ph/9504324.

\bibitem{Care94} M. Carena, M. Olechowski, S. Pokorski and C.E. Wagner, Nucl.\ Phys.\ {\bf B419}, 213 (1994).

\bibitem{Barg97} V. Barger and C. Kao, Phys.\ Rev.\ D {\bf 57}, 3131 (1998).

\bibitem{Abel00} S. Abel {\it et al.}  [SUGRA Working Group Collaboration], hep-ph/0003154.

\bibitem{Abdu99} S. Abdullin {\it et al.}  [SUSY Working Group Collaboration], hep-ph/0005142.

\bibitem{Acco98} E. Accomando {\it et al.}  [ECFA/DESY LC Physics Working Group Collaboration], Phys.\ Rept.\ {\bf 299}, 1 (1998).

\end{thebibliography}
\end{document}